\documentclass[12pt]{article}

\usepackage{graphicx}

\begin{document}
\renewcommand{\thefootnote}{\fnsymbol{footnote}}
\newpage
\pagestyle{empty}
\setcounter{page}{0}

\newcommand{\LAP}{%
  {\small E}%
  {\normalsize N}%
  {\large S}%
  {\Large L}%
  {\large A}%
  {\normalsize P}%
  {\small P}}
\newcommand{\sLAP}{
  {\scriptsize E}%
  {\footnotesize{N}}%
  {\small S}%
  {\normalsize L}%
  {\small A}%
  {\footnotesize{P}}%
  {\scriptsize P}}
\def\logolapin{\hbox{\includegraphics{enslapp.ps}}}
\def\logolight{\hbox{\bf%
    {\large E}%
    {\Large N}%
    {\LARGE S}%
    {\huge L}%
    {\LARGE A}%
    {\Large P}%
    {\large P}}}
\let\logoenslapp=\logolight
%
%
%
\setbox4=\vbox{\hsize=5.2cm
  \begin{center}
    {\bf Groupe d'Annecy\\ 
      \ \\
      Laboratoire d'Annecy-le-Vieux de Physique des Particules}
  \end{center}}
\setbox5=\vbox to \ht4{\vss\logoenslapp\vss}
\setbox6=\vbox to \ht4{\vss\vbox{\hsize=4.2cm
    \begin{center}
      {\bf Groupe de Lyon\\
        \ \\
        Ecole Normale Sup{\'e}rieure de Lyon}
    \end{center}}\vss}
\centerline{\hss\box4\hfill\box5\hfill\box6\hss}
\centerline{\rule{12cm}{.5mm}}

\vspace{20mm}

\begin{center}

{\LARGE {\bf The U.V. Price for Symmetry Non-Restoration}}\\[1cm]

\vspace{10mm}

{\large Jean Orloff\footnote{email: orloff@lapp.in2p3.fr}}\\[.42cm]

{\em Laboratoire de Physique Th{\'e}orique }\LAP\footnote{URA 14-36 
du CNRS, associ{\'e}e {\`a} l'Ecole Normale Sup{\'e}rieure de Lyon et
{\`a} l'Universit{\'e} de Savoie.}\\[.242cm]

\end{center}
\vspace{20mm}

\centerline{ {\bf Abstract}}

\indent

We study the non restoration of symmetries with a local order parameter in
field theory at finite temperature. After giving an interpretation of the
phenomenon, we show that hierarchy problems are a necessary condition for its
realization in renormalizable theories. We then use a large $N$ treatment, and
find that high temperature symmetry can stay broken in this limit (in
opposition with a previous result), and further that the running of couplings
reinforces the effect in the simplest model with two scalars.

\vfill
\rightline{hep-ph/9611398}
\rightline{\LAP---AL---615/96}
\rightline{November 1996}

\newpage
\pagestyle{plain}
\renewcommand{\thefootnote}{\arabic{footnote}}

\section{Introduction}
\label{sec:intro}

Since soft pion theorems in the sixties and renormalization of the standard
model in the seventies, spontaneous symmetry breaking has become an inevitable
building block in particle physics models to understand the small mass of
certain scalar particles, or the large mass of certain gauge ones. It was later
noted \cite{Kirzhnits:1972,Kirzhnits:1975} that in analogy with
ferro-magnetism, heating up such models tends to restore the broken symmetry.
There are two related intuitive arguments why this should be expected. The
first one appeals to thermodynamics: for higher temperatures, the minimum of
the free energy becomes determined less by energy (which has broken symmetry
minima), than by entropy (which intuitively is maximum in symmetric
configurations). Alternatively, one can say that thermal fluctuations are able
to cross the potential barriers surrounding the broken minima, and feel the
global symmetry of the theory.

This implies that any symmetry broken today was once unbroken when the
cosmological temperature was high enough, and there must have been a phase
transition where this freezing of symmetry occurs. During this
freezing, various topological defects can be created. Certain of these, like
monopoles or domain walls are experimentally excluded, which gives constraints
on the particle physics models. For instance, this makes it hard to explain the
breaking of discrete symmetries (like CP) dynamically: at the phase transition
where the symmetry gets spontaneously broken, domain walls separating the
regions with different discrete minima would be formed, and these could quickly
dominate the energy density of the universe.

To relax these constraints, it has been noted \cite{Dvali1:1995}
that domains would not form if the symmetry remains broken, however high the
temperature. This is symmetry non-restoration (SNR).

In one of the founding papers of finite temperature field theory,
Weinberg\cite{Weinberg:1974} already noted that high temperature symmetry
restoration was not an unavoidable fate in field theory, and produced the
following simplest example. Consider a scalar theory with 2 fields in the
vector representations of an $O(N_1)\otimes O(N_2)$ global symmetry group. The
most general renormalizable potential is then:
\begin{equation}
  V(\phi)= {1\over 2}{m_1}^2 {\phi_1}^2  + {1\over2}{m_2}^2 {\phi_2}^2 +
  {1\over4}\lambda_1 {\phi_1}^4 + {1\over4}\lambda_2 {\phi_2}^4
  + {1\over2}\lambda_{12} {\phi_1}^2 {\phi_2}^2,
  \label{eq:pot}
\end{equation}
for which boundedness at the classical level requires
\begin{equation}
  \lambda_1 \lambda_2\ge\lambda_{12}^2; \qquad \lambda_1, \lambda_2\ge 0.
  \label{bound}
\end{equation}
Notice this does not exclude a negative mixed coupling $\lambda_{12}$, provided
it is not excessively large.

For high enough temperatures and small enough couplings, the effective mass
corrections are dominated by the quadratic divergence of the one loop
tadpoles:
\begin{equation}
  m_2^2(T)= m_2^2 + {T^2\over 12} \left[(N_2+2)\lambda_2 + N_1
    \lambda_{12}\right] (1+O(\lambda,m/T))
\label{eq:mevol}
\end{equation}
and similarly for $(1\leftrightarrow2)$. Thus, for
\begin{equation}
  \sqrt{\lambda_1 \lambda_2} \doteq \lambda_{12}^{max} >
  -\lambda_{12} > {N_2+2\over N_1} \lambda_2,
  \label{limits}
\end{equation}
any increase in temperature tends to increase the thermal expectation value of
$\phi_2$
\begin{equation}
  \left<\phi_2\right> = \sqrt{-m_2^2(T)\over\lambda_2} \sim T
\end{equation}
up to arbitrarily large values, while the potential stays bounded from below.

To get an intuitive understanding of this disturbing result, it helps to plot
the equipotentials in the $(\phi_1,\phi_2)$ plane. To simplify, we will
consider the limit of large fields relevant for high temperatures,
and consequently neglect the mass terms in this plot.
\begin{figure}[htbp]
  \begin{center}
    \leavevmode
    \includegraphics{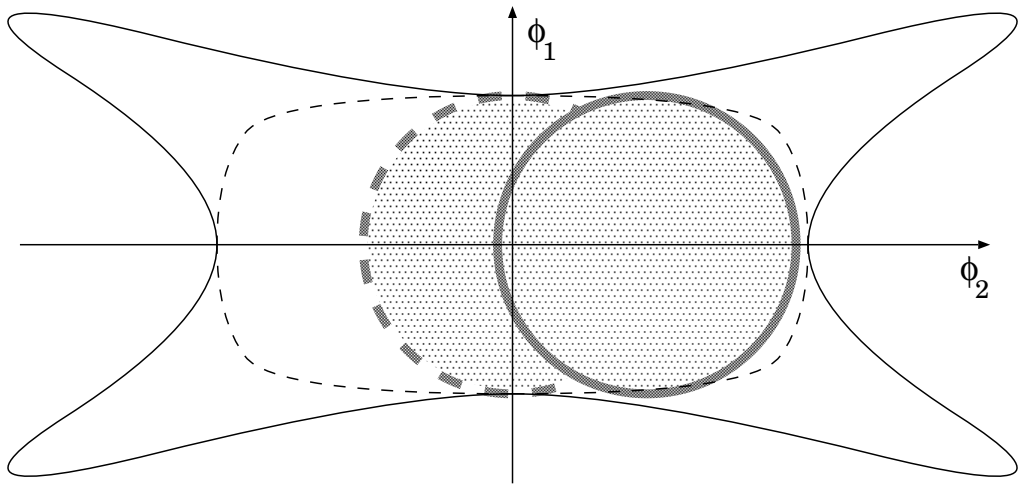}
    \caption{Contour plots of the classical potentials (\ref{eq:pot}),
      for $\lambda_{12}=0$ (thin dotted line) and $\lambda_{12}$ close to 
      $-{\lambda_{12}}^{max}$ (thin plain line). Thick grey disks represent
      thermal fluctuations about the cold (dotted line) or hot (plain line)
      vacuum expectation value.}
    \label{fig:pot}
  \end{center}
\end{figure}
The thin dashed line is an equipotential for the case of vanishing mixed
coupling $\lambda_{12}$. Turning on a negative $\lambda_{12}$, we see the thin
plain equipotential grows diagonal extensions that would eventually make the
potential bottomless for $-\lambda_{12} > \lambda_{12}^{max}$. Since all other
equipotentials are homothetical to the ones drawn, the minimum of this
classical potential lies at the origin. However at finite $T$, we should not
only minimize the potential energy for a point, but also for the thermal
fluctuations around it, roughly depicted as thick grey disks of size $\sim T$.
Because of the pinch in the vertical $\phi_1$ direction, the dashed oval get
pushed away from the origin, to reach the finite $T$ expectation value
surrounded by the plain oval.

With this picture in mind, symmetry non restoration seems to be an indubitable
property of certain scalar theories, at least when couplings are small enough
that one loop calculations can be trusted. It is thus not surprising that
certain crystals (which typically have a rich scalar, vector, tensor, ...
excitation spectrum) undergo such an inverse transition where the higher
temperature lattice has \emph{less} symmetry than the lower $T$ one.  However,
this comparison also calls for a distinction between inverse symmetry breaking
and symmetry non restoration. Indeed, by increasing temperature enough, these
crystals inevitably melt, thus restoring the full spatial group in the liquid
phase. This stresses that any claim for symmetry non restoration is a strong
statement about the ultra-violet structure of the theory.

In section \ref{sec:hier}, we will see that for inverse symmetry breaking to be
possible in a renormalizable theory, its scalar sector must necessarily suffer
from hierarchy problems. This excludes renormalizable supersymmetric theories.

In section \ref{sec:running}, we try to see whether there is any sign for
symmetry restoration when the model (\ref{eq:pot}) is heated close to the point
where its couplings get uncontrollably large (as must somehow happen for
crystal excitation modes close to melting). We find contrarywise that the
running of couplings seems to increase the symmetry non restoration region.

\section{The U.V. Price for Inverse Symmetry Breaking: Hierarchy Problems}
\label{sec:hier}

Consider the most general renormalizable field theory, containing scalars
$\phi_i$, fermions $\psi_\alpha$ and vectors $A_a$. It is easy to prove that if
the theory is hierarchy safe, it cannot undergo inverse transitions. By
``hierarchy safe'' we here mean the naivest possible definition, namely that
relations between dimensionless couplings prevent all dimensional parameters in
the lagrangian from receiving power-like divergencies at one loop. 

Let us in a first step exclude scalar singlets. To decide whether symmetry is
broken or not, we must study the stability of $\langle\phi_i\rangle=0$ which is
determined by the lowest non-trivial terms of the potential. Although scalar
trilinear couplings exist which could give rise to linear terms at one loop,
group theory forces such linear terms to vanish. We must thus focus on mass
terms, which by the above hierarchy requirement only receive logarithmic
radiative corrections. The quadratic divergencies exactly cancel at one loop
between bosons and fermions:
\begin{equation}
  m_i^2= (c_{i,b}-c_{i,f}) {\Lambda_{UV}^2\over 16\pi^2} +
  O(\log \Lambda_{UV});\qquad
  c_{i,b}\equiv c_{i,f},
  \label{hierarchy}
\end{equation}
where $c_{i,b(f)}$ is a combination of all couplings of scalar $i$
to bosons (fermions).  Such a cancellation is for instance automatic in
supersymmetric theories \cite{Mangano:1984}, but the argument here is simpler
and more general.

Turning on temperature, we find
\begin{eqnarray}
  m_i^2(T)-m_i^2(0)&=&c_{i,b} {T^2\over 12} + c_{i,f}
  {T^2\over 24} + O(\log T)\\ 
  &=&3 c_{i,f} {T^2\over 24} + O(\log T),
  \label{thierarchy}
\end{eqnarray}
using ``hierarchy safety'' to obtain the last line. Contrary to bosonic
contributions like (\ref{eq:mevol}), $c_{i,f}$ can easily be checked to
be positive semi-definite. Indeed, they are the eigenvalues of a mass matrix
$g_{ij}$ of the form:
\begin{equation}
  g_{ij}=\sum_{a,b}  y_{iab} y_{jab}^\ast
  \doteq \sum_{A\doteq(a,b)}y_{iA}  y_{jA}^\ast,
\end{equation}
which can be seen as the metric induced on the space of bosonic fields $i$,
from the identity metric $\delta_{AB}$ in the space of fermionic pairs $A$, by
the linear embedding provided by Yukawa couplings $y_{iA}$. By suitable
rotations in the space of bosonic fields, this hermitian matrix can always be
diagonalized:
\begin{equation}
  g_{ij}=\delta_{ij} g_i =\delta_{ij} \sum_{A} |y_{iA}|^2
  \sim\delta_{ij} c_{i,f}.
\end{equation}
This proves that $c_{i,f}$ are positive as claimed, and furthermore, that they
can only vanish for scalars totally decoupled from all fermions. Thus,
$\langle\phi_i\rangle=0$ is a local minimum, whose stability increases as
temperature is raised.

Let us now consider scalar singlets, which allow for new types of linear and
trilinear vertices. The only way these can alter the previous conclusion is by
having some singlets developping a quadratically divergent expectation value
which could contribute to (\ref{hierarchy}) through trilinear coupling.
However, this is forbidden by our naive definition of hierarchy safety as it
induces an even stronger problem than the usual one for scalar masses. One can
for instance show that certain fermions then necessarily receive quadratically
divergent contributions to their mass.
Hierarchy problems in the scalar sector are thus necessary conditions to have
symmetry non-restoration, or even just inverse symmetry breaking. 

\section{Gap Equations and Running Couplings}
\label{sec:running}
Let us now focus back on Weinberg's simple model (\ref{eq:pot}) and consider
the question of coupling running. Neglecting mass terms, the zero temperature
one-loop renormalization of the couplings reads ($dt\doteq d\log
\Lambda_{UV}/16\pi^2$)

\begin{eqnarray}
\label{eq:nrun}
  {d\lambda_1\over dt}&=& (N_1+8) {\lambda_1(t)}^2 + 
  N_2 {\lambda_{12}(t)}^2\nonumber\\
  {d\lambda_2\over dt}&=& (N_2+8) {\lambda_2(t)}^2 + N_1 {\lambda_{12}(t)}^2\\
  {d\lambda_{12}\over dt}
  &=& \lambda_{12}(t) [(N_1+2) \lambda_1(t) + (N_2+2) \lambda_2(t)
  +4\lambda_{12}(t)]\nonumber
\end{eqnarray}

Independently of the sign of $\lambda_{12}$, all couplings thus tend to grow
stronger with increasing $\Lambda_{UV}$, just like for a single self-coupled
scalar. As implied by triviality \cite{Fernandez:1992} in the latter case, the
theory is only defined below $\Lambda_{max}$, some physical maximum UV cut-off,
where the couplings start blowing up.

The presence of such a cut-off requires some care in the definition of symmetry
non-restoration. Indeed, the existence of a maximal energy scale means that
symmetry (non-) restoration can only be probed up to that temperature, and no
easy claim can be done about what happens beyond, where the theory becomes
intractable\footnote{Notice this equally casts doubts on the {\em restoration}
  of the symmetry above that temperature}. Furthermore, an upper cut-off opens
new possibilities: taking for instance a single scalar with
$-m^2(T=0)\gg\lambda{\Lambda_{max}}^2$ will forbid symmetry restoration up to
$T\leq\Lambda_{max}$. This trivial effect is however qualitatively different
from what we have with (\ref{eq:mevol}), where a scalar thermal expectation
value constantly increases with temperature.

We will thus focus our attention on the limit $T\gg m(T=0)$, for which we may
as well take vanishing masses to start with, and will call symmetry non
restoration in the presence of a UV cut-off, the possibility that the symmetry
breaking thermal expectation value shows no tendency to decrease as temperature
is increased all the way up to the cut-off.  In the simplest approach
(\ref{eq:mevol}), this happens provided a condition on the couplings is
satisfied:
\begin{equation}
  \label{eq:naivecc}
  C(\lambda_1,\lambda_2,\lambda_{12}) 
  = (N_2+2)\lambda_2 + N_1 \lambda_{12} 
  < 0.
\end{equation}

However, if couplings are larger, the thermal masses (\ref{eq:mevol}) get
comparable with $T$. To keep properly track of $O(m/T)$ terms, (\ref{eq:mevol})
can be replaced, in the large $(N_1,N_2)$ limit, by self-consistent gap
equations \cite{Fujimoto:1985}, also known as bubble, super-daisy, or cactus
resummation:
\begin{eqnarray}
  \label{eq:cactus}
  x_1^2 \doteq {m_1^2(T)\over T^2} &=& {(N_1+2)\over 12} \lambda_1 f(x_1)
  + {N_2 \lambda_{12}\over 12} f(x_2)\nonumber\\
  x_2^2 \doteq {m_2^2(T)\over T^2} &=& {(N_2+2)\over 12} \lambda_2 f(x_2)
  + {N_1 \lambda_{12}\over 12} f(x_1),
\end{eqnarray}
where
\begin{equation}
  \label{eq:deff}
  f(x)\doteq {6\over\pi^2}\int_0^\infty {p^2 dp\over\sqrt{p^2+x^2}}
  (e^{\sqrt{p^2+x^2}}-1)^{-1}
  \sim 1 - {3\over\pi} x + O(x^2)
\end{equation}
is exponentially decreasing for large $x$. The approximation $f=1$ just gives
back (\ref{eq:mevol}), while expanding $f(x)$ around the origin gives
perturbative corrections for any $N$
\cite{Weinberg:1974,Bimonte1:1996,Amelino-Camelia:1996}. The new region of
symmetry non restoration is now bounded by $x_2=0$ which, upon using
(\ref{eq:cactus}) and dropping $1/N$ corrections, becomes:
\begin{equation}
  \label{eq:cactuscc}
  C(\lambda_1,\lambda_2,\lambda_{12}) =
  \lambda_1 - {\lambda_{12}^2\over\lambda_2}
   + {12\lambda_{12}\over N_2\lambda_2} 
   f^{-1}(-N_2\lambda_2/N_1\lambda_{12})^2 
   <0.
\end{equation}

\begin{figure}[htbp]
  \begin{center}
    \leavevmode
    \includegraphics[width=7cm]{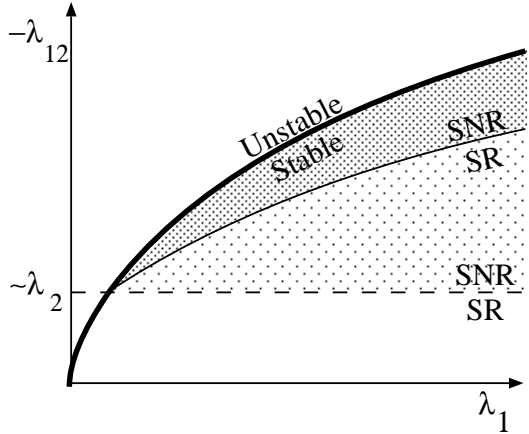}
    \caption{The Symmetry Non Restoration (SNR) domain in couplings space:
      the light grey area is given by naive one-loop calculation; the dark grey
      region is what remains according to gap equations; all stable models
      should lie below the thick curve.}
    \label{fig:snrb}
  \end{center}
\end{figure}

This condition on the couplings contrasts with \cite{Fujimoto:1985}, which
claims the SNR region to be empty in the same large $N$ limit. The origin of
this difference is that we kept all terms in (\ref{eq:cactus}) that stay finite
in the limit $N_{1,2}\rightarrow\infty$ with $\lambda_{1,2,12}N_{1,2}$
finite. One can further explicitly check that there is no $x_2>0$ solution
inside the region (\ref{eq:cactuscc}).

Notice that, just like (\ref{eq:naivecc}), (\ref{eq:cactuscc}) purely gives a
condition on the couplings which are kept constant in the gap equations
approach. Gap equations can not produce a critical temperature because they
have no energy scale other than $T$ in the large temperature limit.

Taking couplings evolution into account will bring in the scale $\Lambda_{max}$
alluded above. To get a feeling for the direction in which this might go, let
us notice that thermal masses come predominantly from ``hard''
momenta of order $T$ in expressions like (\ref{eq:deff}). It is therefore
natural to take the couplings showing up in (\ref{eq:cactus}) as the running
couplings at scale $T$. For a given theory, fixed e.g. by the values of
``bare'' couplings at a scale as close as reasonable to $\Lambda_{max}$, the
running from that initial scale down to $T$ is close to the running we would
have at zero temperature. Indeed, the largest effects of temperature show up in
the infra-red, for scales lower than $T$ where the running effectively becomes
3 dimensional \cite{Tetradis:1993}. We can therefore tentatively use the same
equations (\ref{eq:nrun}) and now interpret $dt$ as $d\log T/16\pi^2$. This
procedure was recently legitimated, at least for small couplings
\cite{Roos:1996}.

It is interesting to notice that the boundary of the stability region
$\lambda_1\lambda_2=\lambda_{12}^2$ is invariant under the large $N$ running of
couplings. This is reassuring about the use of the same condition at all
scales.  Indeed, the global stability of a theory cannot change by integrating
out degrees of freedom. Hence, the finite $N$ result that the flows of
$\lambda$ can cross the ``stability'' boundary must mean that operators higher
than quartic become relevant to this question \cite{Pietroni:1996}.

At every temperature, the symmetry non restoration region is thus given in terms of
a condition on the running couplings at that temperature.  We will now show
that increasing $T$ brings new points in this region, so that symmetry non
restoration is effectively enhanced by couplings running. This region is
bounded by the invariant stability condition and by (\ref{eq:cactuscc}):
\begin{equation}
  \label{eq:l1c}
  \lambda_1=\lambda_{1c}(\lambda_{12},\lambda_2)
  \doteq {\lambda_{12}^2\over\lambda_2}
   - {\lambda_{12}\over N_2\lambda_2} 
   g(-N_2\lambda_2/N_1\lambda_{12})
\end{equation}
where $g(x)\doteq f^{-1}(x)^2$. Let us compare the evolution of $\lambda_1$ as
given by (\ref{eq:nrun}), with the evolution along the surface given by:
\begin{eqnarray}
  {d\lambda_{1c}\over dt} &\doteq&
   {\partial\lambda_{1c}\over\partial\lambda_{12}}
    {d\lambda_{12}\over dt} +
   {\partial\lambda_{1c}\over\partial\lambda_{2}}
    {d\lambda_{2}\over dt}\\
  &=& N_1 \lambda_{1c}^2 + N_2 \lambda_{12}^2 +
      {\lambda_{12}\over N_2 \lambda_2} 
        g' g(-{N_2\lambda_2\over N_1\lambda_{12}})\\
  &=& {d\lambda_1\over dt} +
      {\lambda_{12}\over N_2 \lambda_2} 
        g' g(-{N_2\lambda_2\over N_1\lambda_{12}})\label{eq:piercing}
\end{eqnarray}
The last term of this expression is positive definite, since $\lambda_{12}$ is
negative, and $g$ is purely decreasing ,just as $f$ and $f^{-1}$. As a result,
when increasing $t$, a point on the surface
$\lambda_{1c}(\lambda_{12},\lambda_2)$ is moved to a point with a smaller
$\lambda_1$ than its projection on the surface. This means that the domain of
symmetry non restoration increases with temperature.

Let us now critically examine the approximations used to obtain this result, in
comparison with the exact renormalization group equations used in
\cite{Roos:1996}.  We have first neglected the running of masses as a function
of the momentum scale in the gap equations. Taking this into account would
change the function $f(x)$ in (\ref{eq:deff}), but it could not change its
purely decreasing behavior, which was the only information needed to reach
our conclusion.

We have further neglected masses in the running of couplings (\ref{eq:nrun}).
These would introduce different factors for the loops with different masses.
However, the remarkable cancellations leading to (\ref{eq:piercing}) would
still work, leaving just an $m_1$-depending factor in the last term of that
equation. Since the bare theory needs a large negative $m_1^2$ to achieve a
small infrared $m_1^2$, this factor is dominated in the large $N$ limit by the
$N_1-1$ massless Goldstone modes, for which (\ref{eq:piercing}) is thus
unaltered.

Finally we have kept couplings fixed in the gap equations. This is a good
approximation if couplings evolve slower than masses, like at zero temperature
where their logarithmic running is negligible compared with the power-like mass
evolution. However this assumption clearly breaks down when nearing the
triviality pole. It should be noted that for $T\sim\Lambda_{max}$, we loose
control over the theory anyway, as unavoidable regulator dependencies start
creeping in. This shows up in the exact renormalization group approach
\cite{Roos:1996} under the form of strong oscillations in the $T$ dependence
when $T/\Lambda_{UV}$ is not small. These prevent any firm conclusion in this
region, and the approximation made at least has the merit of cutting them.
Going beyond our approximation requires either knowledge of the theory beyond
$\Lambda_{max}$, or at least a way to reach higher temperatures. This is
possible by introducing a different spatial and temporal cut-off. The spatial
one is still limited by the same $\Lambda_{max}$, but the temporal cut-off can
now be separately sent to infinity, thus allowing for arbitrarily large
temperatures. Physically, this corresponds to measuring temperature by the tiny
energy dependence in the distribution of modes much smaller than $T$.  It has
been proven on the lattice \cite{Bimonte3:1996} that this leads to symmetry
restoration. Although this is far from the continuum limit, we expect a similar
conclusion with a momentum cut-off. 

Outside of this admittedly artificial regularization procedure, the theory can
only be defined beyond $\Lambda_{max}$ by introducing new physics which may
restore the symmetry, or not. The only known fact is that in the latter case,
the hierarchy problems get worse beyond $\Lambda_{max}$.

\section{Conclusion}
\label{sec:concl}

In this paper we studied the non restoration of symmetries (SNR) with local
order parameters. We showed that in the framework of renormalizable theories,
inverse symmetry breaking (and thus a fortiori SNR) necessarily rested on a
hierarchy problem for the scalar field associated with the broken order
parameter. This excludes renormalizable supersymmetric theories. It has
recently been shown \cite{Bajc:1996} that including non-renormalizable
operators does not remove this constraint.

We have then shown that in the simplest model of Weinberg (\ref{eq:pot}), the
inclusion of running couplings does not lead to symmetry restoration, even when
venturing into the ultraviolet strong coupling regime by means of large $N$
techniques.  On the contrary, the running of couplings tends to enlarge the
region of SNR, in accordance with the recent result of
\cite{Amelino-Camelia:1996} on the effect of next to leading order corrections,
the first ones sensitive to coupling running. 

This all should not obscure the fact that this model possesses an intrinsic
maximum energy scale, which translates into a maximum temperature. Such a
maximum scale is quite general in theories with scalars, and is for instance
unavoidable when the scalars fully break all non-Abelian gauge symmetries
\cite{Cheng:1974}.  Without pretending to too much generality, we can say we
failed to construct a model with SNR where all couplings enjoy asymptotic
freedom. The difficulty lies in that the scalar couplings must be large enough
to dominate certain scalar self-energies.

This maximum scale poses a problem for the application of SNR to eliminate
topological defects in cosmology.  Since classical cosmology should be defined
for temperatures up to the Planck scale, we have the following dilemma. Either
we tune parameters so that the maximum scale is pushed above the Planck one. In
that case, it could be argued that the classical universe is born in the broken
phase, but one should somehow cope with graviton loops, and show they do not
alter this statement. Or SNR only holds up to some temperature below the Planck
scale. In that case, inflation is needed to dispose of the topological defects
created at that temperature, and the role of SNR is just to push the moment
where these defects appear back in time, earlier than inflation. However, the
higher in energy we push them up, the more critical the necessary hierarchy
problem becomes. This seems an unavoidable ultra-violet price for SNR.

To conclude, let us stress that we did not touch here symmetry realizations
with a non-local order parameter. For these, the picture is totally different.
For instance, a duality transformation can exchange high and low temperatures.
However, the order parameter broken in the high temperature phase is then
non-local. Similarly, the Polyakov loop is a non-local order parameter for the
deconfining phase transition, and its high temperature thermal expectation
value breaks a center $Z_N$ symmetry, albeit in a controversial
way\cite{Smilga2:1994,Kogan:1992}.

\medskip {\bf Acknowledgements:} It is a pleasure to thank B. Gavela, D.
Jungnickel, M.  Pietroni, N. Rius and C. Wetterich for most informative and
enjoyable discussions. It is not a pleasure to thank the bag thief that got me
to rewrite and simplify the computations.

\bibliography{snr2}
\bibliographystyle{unsrt}
\end{document}